\documentclass[aps, prl, reprint, showkeys, groupedaddress, letterpaper]{revtex4}
\usepackage{epsfig,psfrag}
\usepackage{graphicx,epstopdf}
\usepackage[usenames,dvipsnames]{color}
\usepackage{amssymb,amsmath,amsthm,latexsym}
\usepackage{hyperref}

\def\cF{{\mathcal F}}
\def\cE{{\mathcal E}}
\def\cB{{\mathcal B}}
\def\cC{{\mathcal C}}

\def\cF{{\mathcal F}}
\def\bE{{\mathbf E}}
\def\bq{{\mathbf q}}
\def\bv{{\mathbf v}}
\def\bF{{\mathbf F}}
\def\bJ{{\mathbf J}}
\def\bn{{\mathbf n}}
\def\be{{\mathbf e}}
\def\bj{{\mathbf j}}
\def\bV{{\mathbf V}}
\def\bQ{{\mathbf Q}}
\def\bR{{\mathbf R}}

\def\bD{{\mathbf D}}
\def\cB{{\mathcal B}}
\def\cC{{\mathcal C}}

\def\cA{{\mathcal A}}
\def\bTheta{{\boldsymbol{\Theta}}}

\begin{document}

\title{Speed Distribution of $N$ Particles in the Thermostated Periodic Lorentz
Gas with a Field.}
\date{\today}
\author{F. Bonetto}
\affiliation{School of Mathematics, Georgia Institute of
Technology, Atlanta GA 30332} 
\author{N. Chernov}
\author{A. Korepanov}
\affiliation{Department of Mathematics, University of Alabama at
Birmingham,
Birmingham AL 35294}
\author{J.L. Lebowitz}
\affiliation{Departments of Mathematics and Physics, Rutgers
University, Piscataway NJ 08854} 

\begin{abstract}
 We study the long time evolution and stationary speed distribution of $N$ point
particles in 2D moving under the action of an external field $\bE$, and
undergoing elastic collisions with either a fixed periodic array of convex
scatterers, or with virtual random scatterers. The total kinetic energy of the
$N$ particles is kept fixed by a Gaussian thermostat which induces an
interaction between the particles. We show analytically and numerically that for
weak fields this distribution is universal, i.e. independent of the position or
shape of the obstacles or the nature of the stochastic scattering. Our analysis
is based on the existence of two time scales; the velocity directions become
uniformized in times of order unity while the speeds change only on a time scale
of $O(|\bE|^{-2})$.
\end{abstract}


\maketitle

\bigskip
Our understanding of nonequilibrium stationary states (NESS) of multiparticle
systems is very incomplete at present. In particular, there are no
cases (we know of) where one has an explicit expression for the NESS of an
interacting system of particles with positions and velocities. In this paper we
derive an analytic expression for a non-trivial NESS having a certain
universality. While our proof requires some technical assumptions the arguments
are physically clear and convincing \cite{BCKL2}. The results are furthermore
checked by very extensive computer simulations. 

The model we consider is a variation of the Drude-Lorentz model of electrical
conduction in two dimensions\cite{AM}. The system consists of $N$ particles
(electrons)
moving under the action of a constant external field $\bE$ among a periodic
array of fixed convex scatterers (Sinai billiard) with which they collide
elastically. In order to produce a stationary current carrying state it is
necessary to have a mechanism which will absorb the heat produced by the field
$\bE$. This is modeled here by a Gaussian thermostat which keeps the total
kinetic energy of the system constant \cite{MH}. Thermostated systems are known
to give results which are in accord with observations on real systems, c.f.\ the
Gallavotti-Cohen fluctuation relation \cite{GC,BGG} and the Ruelle results
about
heat conduction \cite{Ru}.

The equations of motion for the system on the unit two
dimensional torus, which corresponds to an infinite system with periodic
scatterers and periodic initial conditions are, taking the mass of each particle
to be one, \cite{BDL,BDLR,BGG}
\begin{equation}\label{eq2}
\left\{
\begin{array}{l}
   \dot\bq_i = \bv_i  \qquad\qquad i=1,\ldots,N \\
   \dot\bv_i = \bF_i = \bE-\frac{\displaystyle \bE\cdot
\bJ}{\displaystyle U}\,\bv_i + \cF_i \end{array}
\right.
\end{equation}
where
\begin{equation}  \label{eq3}
    \bJ=\sum_{i=1}^N \bv_i,\qquad U = \sum_{i=1}^N |\bv_i|^2
\end{equation}
and $\cF_i$ is the ``force'' exerted on the $i$th particle by collisions with
the
fixed scatterers. These collisions only change the direction but not the speed
of the particle. It is easy to see that due to the Gaussian thermostat,
represented by the $(\bE\cdot \bJ)\bv_i/U$ term, the total kinetic energy of the
system is constant, i.e., $\frac{d}{dt}U=0$.

This system was first introduced, for the case $N=1$, by Moran and Hoover
\cite{MH} where it was found numerically that the NESS had a fractal structure.
This was shown rigorously in \cite{CELS}, where it was proven that this system
has a unique singular SRB (Sinai-Ruelle-Bowen) measure for small $\bE$ which
satisfies Ohm's law. The $N=1$ system  was further investigated both
numerically and analytically in \cite{BDL}\cite{BCKL}.

\begin{figure}
\centering
\begin{tabular}{c c c}
Table A & Table B & Table C\cr
\includegraphics[width=0.25\linewidth]{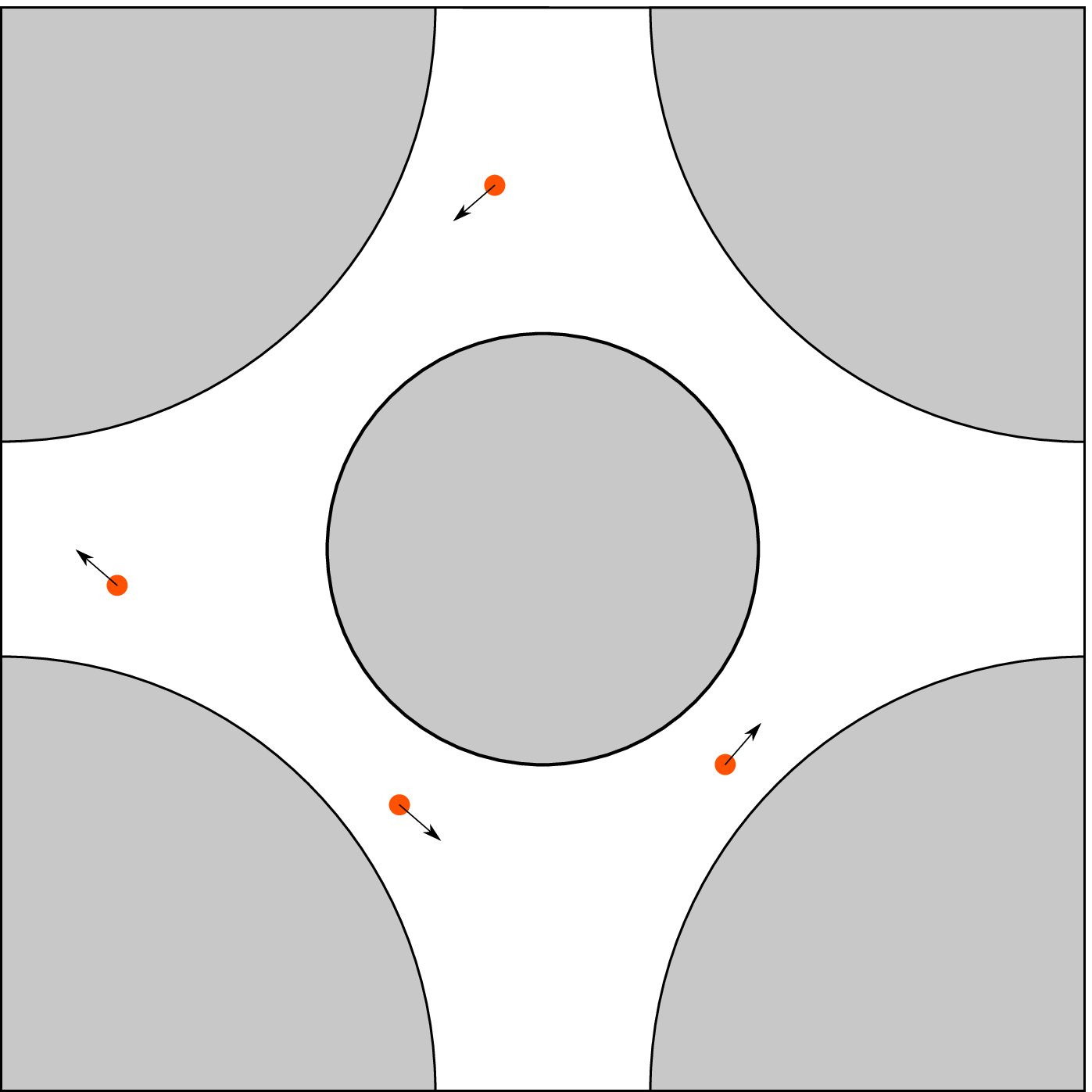}&
\includegraphics[width=0.25\linewidth]{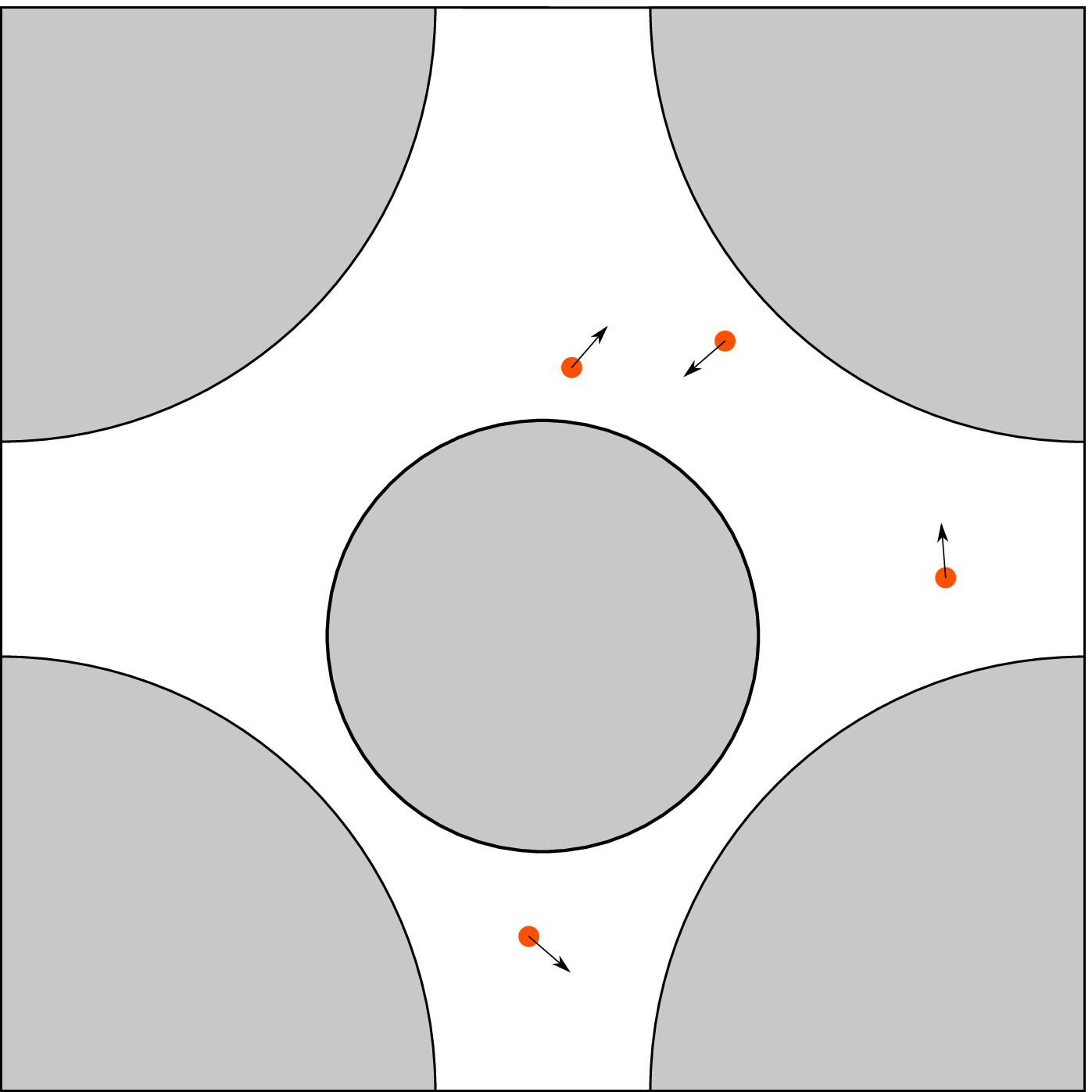}&
\includegraphics[width=0.25\linewidth]{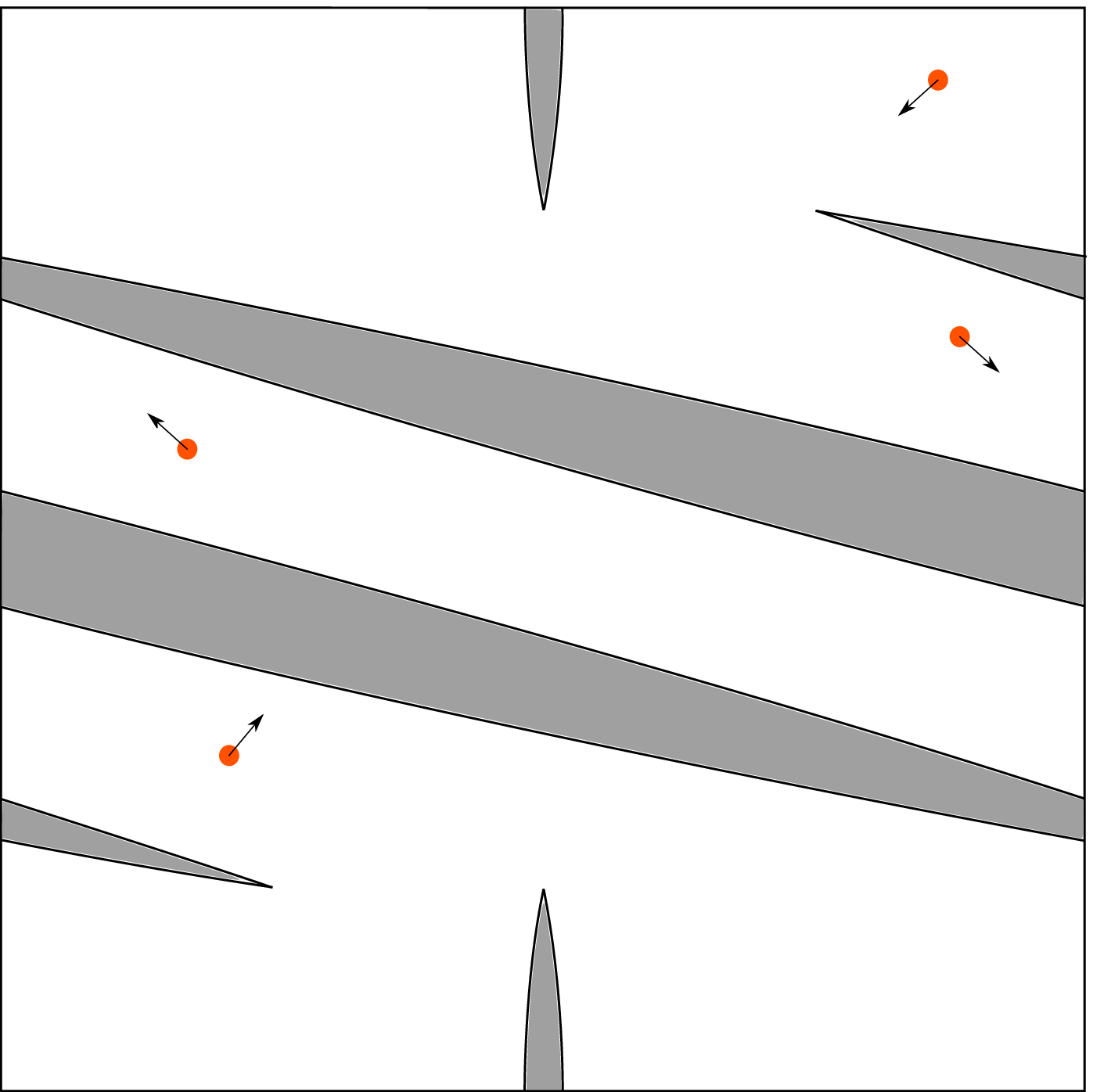}
\end{tabular}
\centering
\caption{Tables used for the simulations.
}
\label{Figure1}
\end{figure}

The multi-particle system $N>1$ was first investigated numerically in
\cite{BDLR}. These gave many tantalizing hints about the interesting nature of
the NESS for this system resulting from the effective interaction between the
particles caused by the thermostat: if one particle increases (decreases) its
speed due to the external field $\bE$ the others have to decrease (increase)
their speed to keep the total kinetic energy fixed. To study the NESS of this
system analytically, a stochastic version of the dynamics was also introduced in
\cite{BDLR}. In this stochastic model the collisions with the fixed obstacles
are replaced by ``virtual'' collisions or scatterings. These collisions, like
the fixed scatterers, conserve energy and tend to make the angular distribution
uniform. It appeared numerically that the one particle marginal speed
distribution of the stochastic and deterministic NESS for a symmetric billiard
table (Table A in Figure \ref{Figure1}) were close for small $\bE$, but there
was no argument of why this should be so. The accuracy of the simulations in
\cite{BDLR} was not very high, so it was assumed that these coincidences were
approximate and valid only for the symmetric table.

We recently revisited this problem using more computer power and new analytical
methods. As a result we now have strong numerical and analytic evidence that the
full NESS speed distribution is, in the limit $\bE\to 0$, exactly the same for
the stochastic and deterministic models for all $N$. This implies ipso facto
that in this limit the distribution is independent of the shape of the billiard
table. It is a ``universal'' function whose exact shape we determine
analytically below. Surprisingly the result seems to remain valid up to
substantial values of $|\bE|$. Just how large $\bE$ can be depends on the shape
of the table; see
Figures \ref{Figure1}. The new element in our analysis is the exploitation of a
time scale separation which occurs for small $|\bE|$.

Going beyond the limit $\bE\to 0$ we find analytically the first order (in
$\bE$) correction to the invariant distribution for a generalized version of
the stochastic model. The expression we find can be extended to the
deterministic billiard where, unlike the speed distribution, it depends on the
shape of the table, but \emph{only}
through some properties which can be obtained from the $N=1$ solution of the
problem. We have checked this expression numerically by computing, with high
precision, the invariant measure for Table B in Figure \ref{Figure1}. We also
find analytically and verify numerically the asymptotic $N>>1$ form of the
speed distribution.
\medskip

\noindent{\bf Analysis:} We shall consider first the stochastic model where the
computations are simpler
and essentially rigorous.
The system now consists of $N$ point particles in the unit 2D torus which move
according to \eqref{eq2} between collisions (without the term $\cF_i$). In
addition each particle independently has a (virtual) collision with a Poisson
rate equal to $\lambda(\bq_i)|\bv_i|$ for some position dependent rate
$\lambda(\bq)> 0$, i.e., the weighted mean free path between collisions
$\int_0^t \lambda(\bq(t))|\dot \bq(t)| dt$ is an exponential random variable
with mean one. The collision changes the angle which $\bv$ makes with the $x$
axis from $\theta'$ to $\theta$ according to some transition kernel
$K(\theta,\theta')\, d\theta$. The exact form of $K$ will turn out not to matter
as long as $K(\theta,\theta') = K(\theta',\theta)$ and there is enough spreading
to the direction of the velocity so that $d\bq\,d\theta/2\pi$ is the unique
invariant distribution for the system with one particle ($N=1$) and $E=0$. The
scattering ``closest'' to that caused by collisions with fixed discs and the one
we used in the simulations is the following: $\bv$ changes to $\bv'$ according
to the rule
\begin{equation} \label{eq4}
   \bv' = \bv - 2\hat{\bn}(\hat{\bn}\cdot \bv)
\end{equation}
where $\hat{\bn}$ is a unit vector in the direction of the momentum transfer
from $\bv$ to $\bv'$. The direction of $\hat{\bn}$ is chosen randomly with
probability density $-(\hat{\bn}\cdot \hat{\bv})/2$, where
$\hat{\bv}=\bv/|\bv|$, subject to the constraint $(\hat{\bn}\cdot \bv)<0$. 

The ``master'' equation describing the time evolution of the $N$-particle
velocity distribution function is, for the above rule, given by
\begin{align}  \label{dW}
 \frac{\partial W(\bQ,\bV,t)}{\partial t} &= 
-\sum_{i=1}^N\bv_i\frac{\partial W(\bQ,\bV,t)}{\partial \bq_i}
-\sum_{i=1}^N
 \frac{\partial}{\partial \bv_i}\Bigl[\bigl(\bE-(\bE\cdot\bj)
\bv_i\bigl)W(\bQ,\bV,t) \Bigr]\nonumber\\
  &\quad +\sum_{i=1}^N\frac 12\int_{(\bv_i'\cdot
  \hat{\bn})<0} \lambda(\bq_i)(\bv_i'\cdot
\hat{\bn})\bigl[W(\bQ,\bV_i',t;\bE)-
W(\bQ,\bV,t;\bE)\bigr]\,d\hat{\bn}
\nonumber\\
   &=\cA W+E\,\cB W+\cC W
\end{align}
where $\bj = \bJ/U$ as  in \eqref{eq3}, 
\begin{equation}
\bQ = (\bq_1,\ldots,\bq_N)\quad\text{,}\quad
 \bV = (\bv_1,\ldots,\bv_i,\ldots,\bv_N)
 \quad\text{and}\quad
 \bV_i' = (\bv_1,\ldots,\bv_i',\ldots,\bv_N),
\end{equation}
and $\bv_i'$ is given in terms of $\bv_i$ by \eqref{eq4}. In the last term 
$E$ is the magnitude of $\bE$, i.e., $\bE=E\be$ for a unit vector $\be$.
Finally, $\cC = \sum_{i=1}^N \cC_i$ and $\cA=\sum_{i=1}^N \cA_i$ are the sum of
collision and streaming terms which act independently and do not depend on $E$.
We shall assume that $\lambda(\bq)$ is such that \eqref{dW} has a unique steady
state solution for every $\bE\not=0$. A sufficient, but not necessary,
condition is that $C^{-1}\leq \lambda(\bq)\leq C$ for some positive $C$. 

Let us consider now what happens when $E$ is small. We note first that
when $E=0$ the speed of any particle does not change with time but the
collisions (deterministic or stochastic) randomize the direction of the velocity
and the position of each particle. The
distribution of speeds would then remain unchanged in time. 
When $E$ is small the appropriate time scale for the change in the speed of
the particle will be of order $E^{-2}$. Now on that time
scale each particle will have undergone many collisions and so one may
then assume that the direction of the velocity and the position of each
particle will be uniformly distributed. 
We can thus expect to have an autonomous equation for the distribution of
the speeds. Let us set
$\bv_i = r_i(\cos(\theta_i),\sin(\theta_i))$ where $r_i = |\bv_i|$ and the angle
$\theta_i$ is
taken with respect to the field direction which we can assume is in the
$x$-direction. Moreover we set $\bR=(r_1,\ldots,r_N)$ and
$\bTheta=(\theta_1,\ldots,\theta_N)$. 
We then carry out a van Hove (weak coupling) limit \cite{Zw,DA},  rescaling of
the time
by letting $t=\tau/E^2$ and set $\widetilde
W(\bQ,\bV,\tau;E)=W(\bQ,\bV,tE^{-2};E)$. $\widetilde
W$ satisfies the rescaled equation

\begin{equation}\label{res}
   \frac{\partial \widetilde W(\bQ,\bV, \tau;\bE)}{\partial \tau}
=E^{-2}(\cA+\cC) \widetilde  W(\bQ,\bV,\tau;\bE) + E^{-1}\cB
\widetilde W(\bQ,\bV, \tau;\bE)
\end{equation}

We now assume that
\begin{equation}\label{eq8}
 \widetilde W(\bQ,\bV,\tau;E)=\widetilde W^{(0)}(\bQ,\bV, \tau)+E
\widetilde W^{(1)}(\bQ,\bV,\tau)+E^2
\widetilde W^{(2)}(\bQ,\bV,\tau)+ o(E^2)
\end{equation}
This is a very strong assumption that can be better justified with a more
detailed analysis \cite{BCKL2}. 

Substituting \eqref{eq8} into \eqref{res} we get the following set of equations

\begin{eqnarray}
 0&=&(\cA+\cC)\widetilde W^{(0)}(\bQ,\bV,\tau)\label{eq24}\\
 0&=&(\cA+\cC)\widetilde W^{(1)}(\bQ,\bV,\tau)+\cB\widetilde
W^{(0)}(\bQ,\bV,\tau)\label{eq25}\\
 \frac{\partial \widetilde W^{(0)}(\bQ,\bV,\tau)}{\partial \tau}
&=&(\cA+\cC)\widetilde W^{(2)}(\bQ,\bV,\tau)+\cB
\widetilde W^{(1)}(\bQ,\bV,\tau)\label{eq26}
\end{eqnarray}
Eq.\eqref{eq24} implies that $\widetilde W^{(0)}(\bQ,\bV,\tau)$ depends only
on $\bR$, i.e., $\widetilde W^{(0)}(\bQ,\bV,\tau)=\widetilde
W^{(0)}(\bR,\tau)$. Since $\cB F(\bV)$ is orthogonal to the functions that
depend
only on
$\bR$ if $F$ depends only on $\bR$, it follows that
$\cE_{\bQ,\bTheta}\cB\widetilde W^{(0)}(\bR, \tau)=0$, where
$\cE_{\bQ,\bTheta}$ is the average on $\bQ$ and $\bTheta$. 
Thank to our hypotheses on $\lambda$ and $K$ we have that
$(\cA+\cC)F(\bQ,\bV)=0$ if and only if $F(\bQ,\bV)$ is a function of $\bR$ alone
so that
$\widetilde W^{(1)}(\bQ,\bV,\tau)=-(\cA+\cC)^{-1}\cB\widetilde
W^{(0)}(\bR, \tau)$ is well defined. We now insert this expression into
\eqref{eq26} and average over $\bTheta$ and $\bQ$. This does not effect the
left hand side since $\widetilde W^{(0)}$ depends only on $\bR$ but it makes
the first term on the right hand side vanish leading to
\begin{equation}\label{BCB}
\frac{\partial\widetilde W^{(0)}(\bR,\tau)}{\partial
\tau}=-\cE_{\bTheta,\bQ}\cB(\cA+\cC)^{-1}\cB
\widetilde W^{(0)}(\bR,\tau).
\end{equation}
which is indeed an autonomous equation for $\widetilde W^{(0)}(\bR,\tau)$. 

Equation \eqref{BCB} can be written out explicitly as:

\begin{equation}  \label{eqF0}
   D^{-1}\frac{\partial \widetilde W^{(0)}(\bR,\tau)}{\partial \tau} =
\sum_{i=1}^N\sum_{j=1}^N
  \frac{\partial^2}{\partial r_i\partial r_j}
  \bigl[M_{ij}(\bR)\widetilde W^{(0)}(\bR,\tau)\bigr] + \sum_{i=1}^N
  \frac{\partial}{\partial r_i}
  \bigl[A_{i}(\bR)\widetilde W^{(0)}(\bR,\tau)\bigr]
\end{equation}
where the component of the $N\times N$ matrix ${\bf M}$ are given by 
\begin{equation}
   M_{ij}(\bR)=\sum_{k=1}^N\frac{b_{ik}(\bR)b_{jk}(\bR)}{r_k}
  =\frac{1}r_i\delta_{ij}-\frac{r_i+r_j}{U}+\frac{r_ir_j}{U^2}\sum_{k=1}^Nr_k\\
\end{equation}
and
\begin{equation}
  A_i(\bR) =
  -\frac{r_i}{U}\sum_{k=1}^N\frac{1}{r_k}+\frac{r_i}{U^2}\sum_{k=1}^Nr_k,
   \qquad b_{ik} = \delta_{ik} - \frac{r_ir_k}{U}
  \label{A_i}
\end{equation}
The diffusion constant $D$ in \eqref{eqF0} is just the
integral of the velocity autocorrelation in the field direction $\be$ when the
magnitude of the field $E=0$, i.e., $D=\be\cdot\bD\be$, where
\begin{equation}
 \bD = \frac{1}{|\bv_1|}\int_0^{\infty} \langle \bv_1\otimes \bv_1(t)\rangle\
  = \frac{1}{|\bv_1|}\int_0^{\infty} \langle \bv_1\otimes
  e^{(\cA_1+\cC_1) t} \bv_1\rangle\, dt,
\end{equation}
and $\langle\cdot \rangle$ is just averaging with respect to the uniform
measure $d\bq\,d\theta/(2\pi)$ that is stationary for $E=0$.

$D$ is in fact the only term in \eqref{eqF0} which depends on the collision
kernel $\cC$ in \eqref{dW}. For a spatially uniform and isotropic scattering,
i.e. when $\lambda(\bq)$ is a constant and
$K(\theta',\theta)=K(\theta'-\theta)$, we get
$\bD=D{\bf I}$ with
\begin{equation}
  D = \frac{1}{2\pi}\int_0^{2\pi} d\theta \int_0^{\infty}
   [cos\theta\cos\theta(t)]\, dt=\frac{1}{2\pi}\int_0^\infty dt \int_0^{2\pi}
d\theta
 \cos(\theta)e^{t\cC_1}\cos(\theta).
\end{equation}
For the specific model used in \eqref{dW}, $D=3/4$. 
In the case of the deterministic billiards $D$ will depend on the shape of the
table. Note
however that the NESS corresponding to the stationary solution of \eqref{eqF0}
is independent of $D$ which really just sets a time scale ($\tau\simeq
t/(DE^2)$). 

We note that
\begin{equation}
 M = SS^{\ast}
 \qquad\text{with}\qquad
 S_{ij}(\bR) = \frac{b_{ij}(\bR)}{\sqrt{r_j}}
\end{equation}
which implies that \eqref{eqF0} corresponds to a stochastic time evolution
described by the It\^{o} stochastic differential equation
\begin{equation}\label{SDE}
 dr_i = -D A_i(\bR)\, dt + \sum_{j=1}^N \sqrt{D} \sqrt{2} S_{ij}(\bR) dB_i 
 \text{,}
\end{equation}
where $B_i$ are $N$ independent Brownian motions.
% \begin{equation}  \label{SDE}
%  \frac{dr_i}{dt} = \sqrt{D}\Bigl[A_i(\bR)+\sum_{j=1}^N
%    S_{ij}(\bR)\,\frac{dr_j}{dt} \Bigr].
% \end{equation}
One can in fact first derive \eqref{SDE} and then obtain
\eqref{eqF0}\cite{BCKL2}. Using
some general theory \cite{Doe} this shows that the solution of \eqref{eqF0} is
unique.

Let now $\widehat W(\bQ,\bV; \bE)$ be the stationary solution of \eqref{dW}. 
Averaging $\widehat W(\bQ,\bV; \bE)$ on $\bQ$ and $\bTheta$ to get $\widehat
W(\bR; \bE)$ and then taking $\lim_{E\to 0}\widehat W(\bR; \bE)$ we get the
stationary solution of
\eqref{eqF0}, $\widehat{W}^{(0)}$. This is in fact the limit as
$\tau\to\infty$ of $\widetilde W^{(0)}(\bR,\tau)$ and is independent of $D$. To
compute it we observe that if $W(\bQ,\bV,t;\bE)$ solves \eqref{dW} so does 
$W'(\bQ,\bV,t;\bE)=h(U)W(\bQ,\bV,t;\bE)$ every positive function $h$. Moreover
\eqref{dW} is invariant under the rescaling
\begin{equation}\label{risca}
 \bV \to \rho \bV,\qquad\qquad t \to \rho^{-1} t,\qquad\qquad\bE \to \rho^2 \bE.
\end{equation}
 This suggests to look for $\widehat{W}_0$ of the form
\begin{equation}
\widehat{W}^{(0)}(\bR)=h(U)F_0(\bR)
\end{equation}
where  $F_0(\rho \bR)=\rho^{2N-1}F_0(\bR)$ and $h(U)$ assures that
$\widehat{W}_0$ has integral 1. With this assumption we get that $F_0$
satisfies the equation
\begin{equation}
 \sum_{i=1}^N \left(\frac{1}{r_i}\frac{\partial^2 F_0}{\partial
r_i^2}+\frac{2}{U}\frac{\partial F_0}{\partial r_i}\right)=0
\end{equation}
This equation can be easily solved and we get, when
the initial state is such that $U=N$,
\begin{equation}  \label{barF0}
 \widehat{W}^{(0)}(\bR)=\frac{1}{Z}\,\delta(U-N)\,\biggl[\sum_{i=1}^N
   r_i^3\biggr]^{-\frac{2N-1}{3}}
\end{equation}
where $Z$ is just the normalization
\begin{equation}
 Z = \int_{\sum r_i^2 = N}\biggl[\sum_{i=1}^N
   r_i^3\biggr]^{-\frac{2N-1}{3}}\,\prod_{i=1}^N r_i\, dr_i
\end{equation}
To get the one particle marginal speed distribution $f_0(r;N)$ one has to
integrate \eqref{barF0} over the variables $r_2,\ldots,r_N$. 
We remark here that when $N\to\infty$ we have 
\begin{equation}\label{infty}
\lim_{N\to\infty}f_0(r;N) = C\exp(-cr^3)
\end{equation}
where 
\begin{equation}
C=\frac{3\Gamma\left(\frac{4}{3}\right)}{2\pi\Gamma\left(\frac{2}{3}\right)^2}
= 0.2325 \qquad\qquad
c=\left(\frac{\Gamma\left(\frac{4}{3}\right)}{\Gamma\left(\frac{2}{3}\right)}
\right)^\frac{3}{2}=0.5355
\end{equation}

Going beyond the limit $E\to 0$ we find the first order correction (in $E$) to
the stationary solution of \eqref{dW}:
\begin{equation}\label{expa}
  \widehat{W}(\bR,\bTheta; E,N) = \widehat{W}^{(0)}(\bR)
  +E \widehat{W}^{(1)}(\bR,\bTheta) + o(E),
\end{equation}
where
\begin{equation}\label{o1}
  \widehat W^{(1)}(\bR,\bTheta; N)=(\cA+\cC)^{-1}\cB\widehat
W^{(0)}(\bR;N)= F_1(\bR)\sum_{i=1}^N
r_ic(\bq_i,\theta_i)
\end{equation}
with $ F_1(\bR)=(2N-1)\biggl[\sum_{i=1}^N
r_i^3\biggr]^{-\frac{2N+2}{3}}$
and
\begin{equation}
  c(\bq_i,\theta_i) = \int_0^\infty
e^{t(\cA_i+\cC_i)}\cos\theta_idt=-(\cA_i+\cC_i)^{-1}\cos\theta_i
\end{equation}
$\cC_i$ and $\cA_i$ are the collision and streaming operators defined on the
right hand side of \eqref{dW}. Note that, for $N=1$, the invariant solution to
\eqref{dW} is simply
\begin{equation}
\widehat{W}(r,\theta; E,1)=\frac{1}{2\pi r}+\frac{E}{r^3}c(\bq,\theta)+o(E)
\end{equation}
so that $c(\bq,\theta)$ is simply related to the $N=1$ problem. 

\medskip
\noindent{\bf Deterministic Billiard:} The master (Liouville) equation for the
deterministic model is given by
\begin{equation}\label{dWd}
  \frac{\partial W_d(\bQ,\bV,t)}{\partial t} =\cA W_d+E\,\cB W_d +\cC_d W_d 
\end{equation}
with $\cC_d W_d$ representing the collisions with the fixed convex obstacles.
In this case the stationary state is not
absolutely continuous with respect to the Lebesgue measure, i.e. it will not
have a smooth density $\widehat W_d(\bQ,\bV;\bE)$, see \cite{CELS,BGG}. On the
other hand it will
have a smooth density $W_d(\bQ,\bV,t;\bE)$ for finite time $t$ and we can
consider the measure 
\begin{equation}
\mu^t_{\bE,N}(d\bQ,d\bV)=W_d(\bQ,\bV,t;\bE,N)d\bQ\,d\bV
\end{equation}
For $N=1$ we know that $\hat \mu_{E,1}(d\bq,d\bv)=
\lim_{t\to\infty}\mu^t_{E,1}(d\bq,d\bv)$ exists and its projection on the
energy surface $|\bv|=1$, can be
written as $\hat\mu_{E,1}(d\bq,d\theta)=d\bq\,d\theta/(2\pi) +
E\delta\mu_1(d\bq,d\theta)+o(E)$. We note that the property \eqref{risca} remain
true for any solution $W_d(\bQ,\bV,t;\bE)$ of \eqref{dWd}.
The expansion \eqref{expa} can be generalized to the deterministic
billiard by replacing $c(\bq,\theta)$ with $\delta\mu_1(d\bq,d\theta)$. More
precisely we have
\begin{equation}\label{expad}
 \hat\mu_{\bE,N}(d\bQ,d\bV)= \delta(U-N)\left(F_0(\bR)\tilde d\bQ\,d\bV+E
F_1(\bR)\sum_i r_i^2
\delta\mu_1(d\bq_i,d\theta_i)\tilde d\bQ^i\,d\bV^i+o(E)\right)
\end{equation}
where  $\tilde d\bQ=\prod_{i}\tilde d\bq_i$,
with $\tilde d\bq$ the normalized restriction of the Lebesgue measure to
$\mathbb{T}^2\backslash$obstacles, $d\bQ^i=\prod_{j\not =i}\tilde
d\bq_j$, $d\bV^i=\prod_{j\not = i}d\bv_j$ and $F_1(\bR)$ is defined after
\eqref{o1}. The above expression implies that
$\hat\mu_{N}(d\bQ,d\bV)=\lim_{\bE\to 0}\hat\mu_{\bE,N}(d\bQ,d\bV)$ is
absolutely continuous with respect to the Lebesgue measure on the energy
sphere and depends only on the speeds. We cannot prove this statement rigorously
but it is well verified by
our numerical simulations involving the full billiard table or just a portion of
it.

\medskip
\noindent{\bf Numerical results:}
We have concentrated our numerical simulation on the deterministic billiard
system using the billiard tables depicted in Figure \ref{Figure1}. Table A is
the same used in \cite{BDLR}. Table B has the central obstacle moved down to
break the symmetry but remaining rather close to Table A. Table C instead was
chosen to be as asymmetric and as far from Table A as possible.

We computed the one particle marginal of the speed distribution for all 3
tables. To obtain an accurate and reliable result we ran a very long trajectory
recording the speed of particle 1 every time of the order of $E^{-2}$. In this
way we can assume that the data we collected form a random sample from the
distribution $f_{0,d}(r,N)$. This allows us to use the Kolmogorov-Smirnov test
to check whether $f_{0,d}(r,N)=f_0(r,N)$ described after \eqref{barF0}, see
\cite{HM,EDJRS}.

\begin{figure}
\centering
\includegraphics[width=0.7\linewidth]{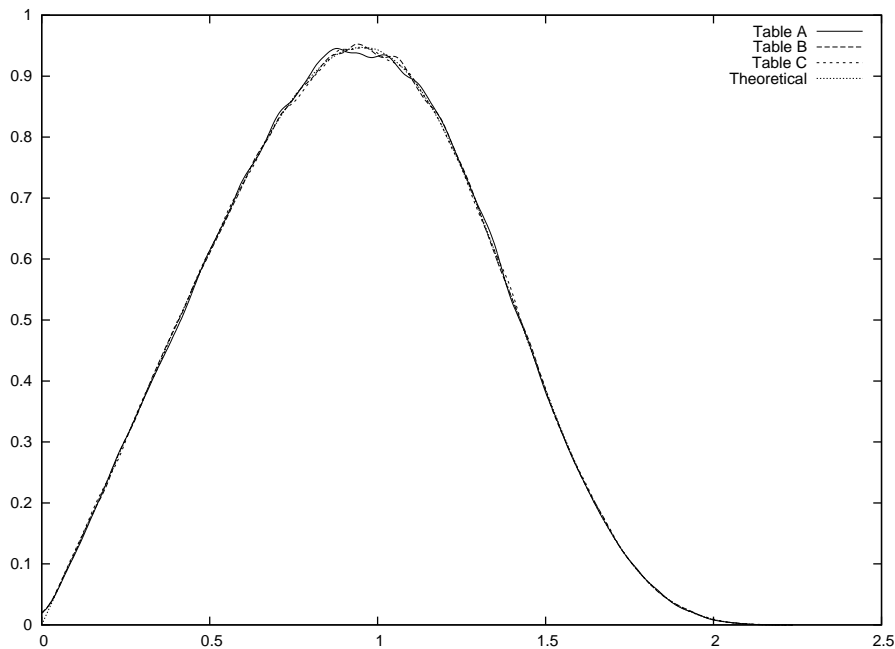}
\centering
\caption{Comparison between the 1 particle speed marginals of the different
tables with 5 particles.}
\label{Figure1A}
\end{figure}

\begin{figure}
\centering
\includegraphics[width=0.7\linewidth]{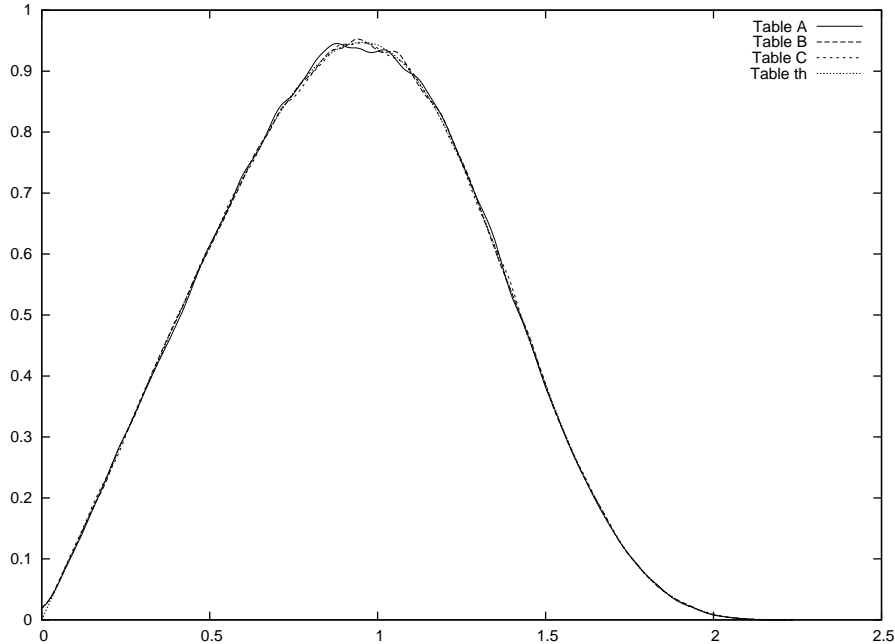}
\centering
\caption{Comparison between the 1 particle speed marginals of the different
tables with 512 particle.}
\label{Figure2A}
\end{figure}

In figure \ref{Figure1A} we plot the marginal distribution for all three table
when
$\bE=0.015625(\cos(\phi),\sin(\phi))$ with $\phi=\pi/2$ and $N=5$ together
with the theoretical prediction coming from \eqref{barF0}. In figure
\ref{Figure2A} we plot
the similar results for $N=512$ and compare it with \eqref{infty}. The
P-value of the KS test for the cases shown in these figures was greater than
$23\%$ giving a strong evidence that our hypothesis on the distribution of the
observed data is indeed correct.A more
extensive report on these simulations can be found at
\url{http://www.math.uab.edu/~khu/g/gt/speed.html}

To check whether the full distribution in \eqref{expad} is valid for
the deterministic billiard we chose $N=3$ and fixed $U=3$ so that we can take
$r_3^2=3-r_1^2-r_2^2$. Fixing $\delta_r=\sqrt{3}/30$, let
$\Delta_{i,j}(r_1,r_2)$ be the characteristic function of the square of side
$\delta_r$ centered at $((i+0.5)\delta_r,(j+0.5)\delta_r)$. We ran 
1000 trajectories of average length $2\cdot 10^8$ system time units for Table B
with $\bE=0.04(\cos(\phi),\sin(\phi))$ with $\phi=\pi/3$. Since it is very
hard to collect enough statistic to properly resolve the full distribution, we
did not attempt to decorrelate the the data nor to run any statistical test on
the reliability of the following results.

We used the above described trajectories to compute
\[
G_c(i,j;k)=\hat \mu_{\bE,3}(\Delta_{i,j}(r_1,r_2)\cos(k\theta_1))\qquad
G_s(i,j;k)=\hat \mu_{\bE,3}(\Delta_{i,j}(r_1,r_2)\sin(k\theta_1))
\]
and
\[
L_c(i,j;k)=\hat
\mu_{\bE,3}(\Delta_{i,j}(r_1,r_2)\cos(k\theta_1)\chi_A(\bq_1))\qquad
L_s(i,j;k)=\hat \mu_{\bE,3}(\Delta_{i,j}(r_1,r_2)\sin(k\theta_1)\chi_A(\bq_1))
\]
with $k=1,2,4,8$. Here $\chi_A(\bq)$ is the characteristic function of the
lower quarter of the table, i.e. $\chi_A(\bq)=1$ if $\bq_x<0.5$ and $\bq_y<0.5$
and 0 otherwise.
Finally the we ran 200 trajectories of length almost $10^9$ system unit time for
an identical system with $N=1$ to compute $g_c(k)=\delta\mu_1(\cos(k\theta))$
and $l_c(k)=\delta\mu_1(\cos(k\theta)\chi_A(\bq))$ and the relative
sine terms. In this way we can check
\eqref{expad} with no parameters to be fitted. Observe moreover that from
\eqref{expad} we have
\[
 G_c(i,j;k)=a_c(k)L_c(i,j;k)
\]
with $a_c(k)=g_c(k)/l_c(k)$, and similarly for $G_s$. In figure 2 we plot
$G_s(i,20;1)$ with $a_s(1)L(i,20;1)$ and the theoretical prediction. We have
selected to plot only $j=20$, corresponding to $r_2=1.184$, for better
readability. On the horizontal axis we have $r_1$ instead of $i$ for better
readability. As one can see, the agreement is very good. Similar agreement is
found for the other $k$. The interested reader may
visit \url{http://www.math.uab.edu/~khu/g/gt/speed.html}
where more plots of the results of these simulations can
be found.

\begin{figure}
\centering
\includegraphics[width=0.7\linewidth]{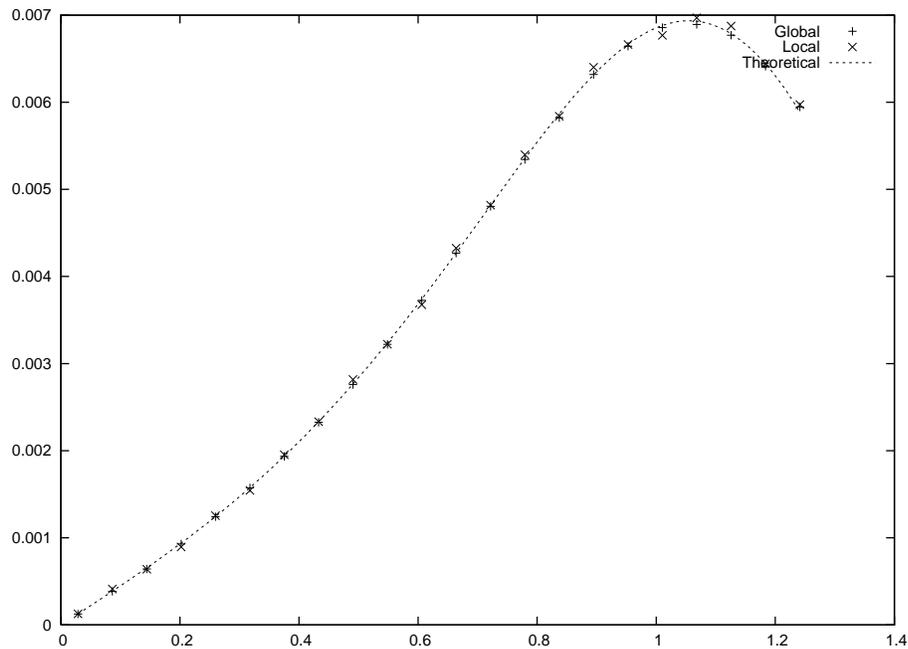}
\centering
\caption{Global and local distributions compared with the theoretical
prediction: $k=1$ sine component.}
\label{Figure2}
\end{figure}

\medskip
\noindent{\bf Acknowledgements.}
The authors thank Michael Loss and Stefano Olla for many insightful comments and
discussions. The work of FB was supported in part by NSF grant 0604518. The work
of NC was supported in part by NSF grant DMS-0969187. The work of JLL
was supported in part by NSF Grant DMR-1104501 and AFOSR grant FA9550-10-
1-0131. The authors are also grateful to the Alabama supercomputer
administration and the Georgia Tech PACE administration for computational
resources.

\end{document}